\documentclass[sigconf]{acmart}

\AtBeginDocument{%
  \providecommand\BibTeX{{%
    \normalfont B\kern-0.5em{\scshape i\kern-0.25em b}\kern-0.8em\TeX}}}

\setcopyright{acmcopyright}
\copyrightyear{2020}
\acmYear{2020}
\acmDOI{10.1145/1122445.1122456}

\acmConference[CIKM '20]{CIKM '20: ACM International Conference on Information and Knowledge Management}{October 19--23, 2020}{Galway, Ireland}
\acmBooktitle{CIKM '20: ACM International Conference on Information and Knowledge Management,
  October 19--23, 2020, Galway, Ireland}
\acmPrice{15.00}
\acmISBN{978-1-4503-XXXX-X/18/06}

\usepackage{multirow}
\usepackage{subfiles}
\usepackage{subcaption}

\begin{document}

\title{\texttt{TRIER}: Template-Guided Neural Networks for Robust and Interpretable Sleep Stage Identification from EEG Recordings}

\author{Taeheon Lee}
\authornote{Both authors contributed equally to this research.}
\authornote{Corresponding author}
\orcid{0000-0003-3777-1872}
\affiliation{%
  \institution{Looxid Labs}
  \city{Seoul}
  \country{Republic of Korea}
}
\email{taeheon.lee@looxidlabs.com}

\author{Jeonghwan Hwang}
\authornotemark[1]
\orcid{0000-0003-4399-1556}
\affiliation{%
  \institution{Looxid Labs}
  \city{Seoul}
  \country{Republic of Korea}
}
\email{jeonghwan.hwang@looxidlabs.com}

\author{Honggu Lee}
\affiliation{%
 \institution{Looxid Labs}
 \city{Seoul}
 \country{Republic of Korea}
}
\email{honggu.lee@looxidlabs.com}

\renewcommand{\shortauthors}{Lee et al.}

\begin{abstract}
  Neural networks often obtain sub-optimal representations during training, which degrade robustness as well as classification performances. This is a severe problem in applying deep learning to bio-medical domains, since models are vulnerable to being harmed by irregularities and scarcities in data. In this study, we propose a pre-training technique that handles this challenge in sleep staging tasks. Inspired by conventional methods that experienced physicians have used to classify sleep states from the existence of characteristic waveform shapes, or template patterns, our method introduces a cosine similarity based convolutional neural network to extract representative waveforms from training data. Afterwards, these features guide a model to construct representations based on template patterns. Through extensive experiments, we demonstrated that guiding a neural network with template patterns is an effective approach for sleep staging, since (1) classification performances are significantly enhanced and (2) robustness in several aspects are improved. Last but not least, interpretations on models showed that notable features exploited by trained experts are correctly addressed during prediction in the proposed method.
\end{abstract}



\maketitle

\section{Introduction}
\label{sec: intro}
Identifying periodic changes in sleep stages is an important task for sleep disorder comprehension \cite{terzano2002atlas,vsuvsmakova2004human}, since clinical diagnosis on patients are inferred from sleep stages. Physiological activities, which are in form of electrical signals gathered via bio-medical sensors like electroencephalography (EEG) and electrooculography (EOG), are inspected in order to map each sleep period to a set of meaningful sleep stages \cite{supratak2017Deepsleepnet,perslev2019u}. This process can be accomplished by trained physicians with expertise and manual annotation of every single sleep period from biomedical signals requires huge amount of time. Hence, a number of researches have been proposed to automate this process. Among these, deep learning techniques have shown promising results, where notable achievements were made from models with convolutional neural networks (CNN) \cite{perslev2019u} and recurrent neural networks (RNN) \cite{supratak2017Deepsleepnet}. These approaches have led recent improvements in sleep staging with state-of-the-art performances achieved in an end-to-end manner.

\begin{table}[b]
  \caption{Sleep Stages and Template Patterns}
  \label{tab:sleep_stages}
  \begin{tabular}{@{}c p{7.2cm}@{}}
    \toprule
    Stage & Description \cite{vsuvsmakova2004human,malhotra2013sleep} \\
    \midrule
    W  & Wakefulness. Patterns characterized by high frequency waves are observed: \underline{alpha} (8-13Hz) and \underline{beta} (13-25Hz).\\
    \rule{0pt}{3ex}
    N1 & Short, transitional phase that occurs when sleep begins. Low-voltage, fast EEG activities are observed: \underline{theta} (4-7Hz) waves. Sometimes, sharp waves are generated. \\
    \rule{0pt}{3ex}
    N2 & Intermediate sleep stage which accounts for 50\% of sleep. Theta waves and quick bursts of neuronal activities occur: \underline{Sleep Spindles} (12-15Hz) and \underline{K-complexes}. \\
    \rule{0pt}{3ex}
    N3 & Deep sleep phase. Extremely slow waveforms with high amplitudes are generated: \underline{delta} (1-4Hz) or slower waves.\\
    \rule{0pt}{3ex}
    REM & Phase where brains become active. Mixed signals of \underline{alpha} and \underline{theta} waves are observed.\\
  \bottomrule
\end{tabular}
\end{table}

Despite these successes, however, there remain important challenges regarding deep learning techniques. Recent studies reported that neural networks often acquire sub-optimal representations during training \cite{brendel2018bagnets,geirhos2018imagenet}. In addition, neural networks are more likely to fail in achieving generalizable representations when data is scarce or irregular \cite{laboy2019normalized}. These challenges become more severe in bio-medical domains like sleep staging where data acquisition process is time-consuming and costly. To overcome these challenges, a number of works were proposed with components that properly guide neural networks to obtain desirable representations. Especially, integration of domain knowledge into deep learning has attracted many researchers \cite{ma2018kame,hong2019mina,li2019domain}. These works successfully improved performances of neural networks on bio-medical domains by introducing proper guidance derived from domain knowledges. These previous works suggest that making a neural network focus on significant features in data could be effective.

This work is proposed with similar regards. Based on conventions that physicians have relied on occurrences and relations between characteristic waveform shapes when annotating EEG recordings for sleep stage scoring \cite{vsuvsmakova2004human}, the proposed work extract such distinctive patterns. Then, these characteristic features are used in pre-training the first convolutional layer of neural networks in order to enable neural networks to fully employ distinctive patterns in recordings. In this paper, we refer these patterns as \textbf{Template patterns}. However, defining the exact numerical forms of such signals are not available at the moment, since characteristics of EEG recordings vary significantly across datasets depending on recording channels and patients/experimental conditions. Thus, novel methodology that automatically extracts templates from given datasets should be considered.

In this paper, we propose \textbf{TRIER}, \underline{\textbf{T}}emplate-guided neural networks for \underline{\textbf{R}}obust and \underline{\textbf{I}}nterpretable Sleep Staging Identification from \underline{\textbf{E}}EG \underline{\textbf{R}}ecordings. TRIER is proposed to extract template patterns in recordings and to exploit them during prediction. Here, in order to locate and extract template patterns, we propose a novel convolutional layer. Features in this layer are computed using cosine similarity \cite{luo2018cosine} in place of dot product to correctly reflect waveform shapes regardless of amplitude. Learned filters from the layers are used as pre-trained templates on target neural networks. With the proposed approach, models are capable of obtaining representations built on template patterns, which might be much more generalizable for unseen data. Through extensive experiments, we demonstrated that the proposed method improved baseline architectures in several aspects. First, classification performances are significantly enhanced on several sleep staging datasets. Second, models become \textbf{robust} in various aspects: confidence calibration and robustness to dataset sizes and gaussian noises. Previous works found that improvements in robustness are one of the main earnings that pre-training can bring \cite{hendrycks19a} and our results on robustness indicate that the proposed method correctly obtained these utilities from pre-training. Furthermore, we believe such improvements in robustness are important in fostering neural networks to real-world sleep staging tasks. Finally, with interpretation on learned filters and model behaviors, we empirically show that models trained with the proposed method correctly utilized important features in EEG recordings. Such inspection demonstrates that TRIER is much more \textbf{interpretable} for real-world application, since model successfully revealed significant features referenced by experts.

\section{Related Works}
\label{sec: related_works}
\subsection{Sleep Staging Algorithms}
Previous studies for developing automated sleep staging algorithms counted on manually curated features from EEG recordings. Hand-crafted features were mainly adopted to construct meaningful representations. For instance, specific form of wavelet transformation, which exploits energy level of well-known frequency bands of EEG recordings, was introduced to build effective representations for classification \cite{oropesa1999sleep}. Another work integrated time-frequency domain information into non-linear features derived from prior knowledge in EEG signals \cite{koley2012ensemble}. Novel form of filters in corporation with previously investigated statistical features in EEG recordings were proposed to extract features from the designated frequency bands \cite{aboalayon2014efficient}. These approaches properly constructed representations for EEG recordings based on domain knowledge on the signal. However, they have major shortcomings that they require manual engineering of parameters to find optimal configurations. Recent approaches on sleep staging employed deep learning model. With deep learning, features were constructed automatically with neural networks in an end-to-end manner. Furthermore, these models surpassed the performances of the manually curated models. In these models, convolutional layers are generally employed in feature extracting layers to learn local patterns in data \cite{sors2018convolutional,perslev2019u}. Some works further exploited recurrent neural networks after the convolutional feature extractors to model temporal relations between sleep stages \cite{supratak2017Deepsleepnet,mousavi2019sleepeegnet}. Unlike these works where deep learning is just employed in an end-to-end manner, our work further improved baseline architectures by properly guiding models with desired features.

\subsection{Confidence Calibration}
Besides classification accuracies, several aspects of neural networks arouse as important research problems. A number of recent works pointed out that neural networks are tend to behave in undesirable ways. First of all, modern neural networks are known to be over-confident even for wrong predictions \cite{guo2017calibration,lakshminarayanan2017simple}. This is a severe problem in mass adoption of deep learning, since people cannot reliably trust the outputs of models. Second, neural networks are not robust against small changes in inputs. Recent works reported that trained models are usually prone to produce wrong predictions even for imperceptible perturbations \cite{szegedy2013intriguing,dodge2017study}. Considering irregularities in data and consequences of model predictions, these vulnerabilities should be taken into consideration when applying neural networks to bio-medical domains. Researchers are starting to work on model robustness in these perspectives. In the analysis of electrocardiogram (ECG), MINA \cite{hong2019mina} addressed model's robustness against distortion and perturbations on signals. For sleep staging, REST \cite{duggal2020rest} introduced novel loss terms for robustness against various forms of noises. Our work also considers robustness of models on gaussian noises similar to these works. However, this is the first work to bring confidence calibration issues to sleep staging tasks.

\subsection{Guidance on Neural Network Training}
Recent findings pointed out that neural networks could acquire undesirable representations, which do not work out robustly. For example, classifiers on image datasets rely heavily on biases from textures other than shape information, which is against our intuition \cite{brendel2018bagnets,geirhos2018imagenet}. To cope with these limitations, various techniques, which are based on novel data synthesis \cite{geirhos2018imagenet} or representation de-correlating techniques \cite{wang2019learning}, were proposed to guide models to learn desirable representations from human's perspectives. For analysis of time-series waveforms, representations from the first layer of convolutional layers are often tackled. SincNet \cite{ravanelli2018speaker} parameterized the first convolutional layer to a sinc function that is designed to operate as a band-pass filter. Wavelet decomposition network defined the first layer to work as wavelet decomposition when analyzing time-series data \cite{wang2018multilevel}. Our work is similar to these works as we configured the first convolutional layer with template signals.

In bio-medical domains, proper representations are often derived from domain knowledge. Previous study employed attention mechanism to integrate beat information from three different levels, which are referenced by medical knowledge in analyzing ECG signals \cite{hong2019mina}. Other work introduced auxiliary task of locating key-features like P-peaks and R-peaks during training in order to guide neural networks \cite{li2019domain}. These works further utilized these features when visualizing evidences used during prediction. To the extent of our knowledge, this is the first work to introduce proper guidance to models, which performs sleep staging in the way that real-world experts are annotating EEG recordings.

\section{Methods}
\label{sec: proposed_approach}
Utilizing prior knowledge from domain fields might be helpful as shown by a number of previous works \cite{hong2019mina,li2019domain}. In this work, we provide target neural networks with prior knowledge that experts used to locate specific waveform patterns in EEG recordings and inspect relations between them. Proposed strategy is to pre-train the first convolutional filters with such template patterns; thus significant parts in data are first recognized. Afterwards, overall representations are constructed based on them. During pre-training phase, we introduce a novel form of neural network in which cosine similarity based convolutional layer and 1-max pooling architecture are deployed in order to effectively locate template patterns from data. Notations for defining the method is listed in Table \ref{tab:notations}.

\begin{table}[b]
  \caption{Notations for Describing the Proposed Approach.}
  \setlength{\tabcolsep}{4pt}
  \renewcommand{\arraystretch}{1.1}
  \centering
  \begin{tabular}{@{}l@{\hskip 0.1in} l@{}}
    \toprule
    $\mathbf{x}$                    & Input to the neural network (EEG recording)                                             \\
    $\mathbf{w}$                    & Weight vector of a convolutional filter                                                 \\
    $\mathbf{o}$                    & Output vector from a convolutional filter                                               \\
    $\mathbf{w}^k$                  & $k^{th}$ filter vector of a convolutional layer                                         \\
    $\mathbf{o}^k$                  & $k^{th}$ channel output vector from a convolutional layer                               \\
    \rule{0pt}{3ex}$E$              & Loss value calculated from the prediction and label                                     \\
    $L$                             & Length of a convolutional filter                                                        \\
    $i^{*}$, $j^{*}$                & \begin{tabular}[t]{@{}l@{}} Index of a convolutional output vector that has             \\
                                                               maximum activation value \end{tabular}                         \\
    \rule{0pt}{3ex}$v_i$            & Scalar value at the $i^{th}$ index of a vector $\mathbf{v}$                             \\
    $\mathbf{v}_{[i:j]}$            & Segment of a vector $\mathbf{v}$ from index $i$ to $j$                                  \\
    $\vert \mathbf{v} \vert$        & L2-norm of a a vector $\mathbf{v}$                                                      \\
    \bottomrule
  \end{tabular}
  \label{tab:notations}
\end{table}

\subsection{1-max Pooling Layer Architecture for Extracting Significant Patterns}
Learned filters in the first convolutional layer of neural networks often lack distinguishable features when analyzing raw time-series recordings. Previous studies reported seemingly noise-like shapes in the learned filters \cite{ravanelli2018speaker,lee2018samplecnn}. We believe such representations might not be template patterns in data. Thus, a novel approach is proposed.

\textbf{Motivation}. In general 1D convolutional layer, derivatives with regard to a weight vector $\mathbf{w}$ can be written as 
\begin{equation}
    \frac{\partial E}{\partial \mathbf{w}} = \sum_{i=1}^{l_{out}}{\frac{\partial E}{\partial o_i}\mathbf{x}_{[i;i+l-1]}},
\end{equation}
with weight values after update
\begin{equation} \label{eq:weight_update}
    \mathbf{w}^{new} = \mathbf{w}^{old} - \alpha \sum_{i=1}^{l_{out}}{\frac{\partial E}{\partial o_i}\mathbf{x}_{[i;i+l-1]}}.
\end{equation}
Applying Fourier Transform to each side of equation \ref{eq:weight_update}, we can decompose the gradient value in terms of frequency components. By the linearity of Fourier Transform, weight updating process becomes identical to adding frequency components of each input segment to current filters, weighted by gradient values. As a consequence, resulting summation becomes noisy, since total gradient values might be intermingled by various frequency components, leading to indistinguishable shapes. 

\textbf{Proposed Approach}. We exploited 1-max pooling layer to construct distinctive template representations on each of convolutional filters. With 1-max pooling layer, gradients regarding output values, $\frac{\partial E}{\partial o_i}$, become zero except for the index $i^{*}$ with maximum output value. Now, weight update equation will be simplified as:
\begin{equation}
    \mathbf{w}^{new} = \mathbf{w}^{old} - \alpha \frac{\partial E}{\partial o_{i^{*}}}\mathbf{x}_{i^{*}+l-1}.
\end{equation}
Here, only the waveform patterns in the input segment that is regarded as the most important patterns will be reflected to parameter update during training. Actually, convolutional layer with one-max pooling has already been proposed to learn significant patterns in data. For DNA sequences and protein sequences, convolution with 1-max pooling successfully captured significant patterns in input sequences \cite{alipanahi2015predicting,seo2018deepfam}.

We employed 1-max pooling architecture in a similar manner to extract template patterns. However, in case of bio-medical recordings, amplitudes of recordings vary a lot even within the same recording. Affected by amplitudes of input values, normal convolutional operator might not find the location of important features properly. For this reason, additional approach that can effectively handle amplitude changes should be introduced.

\subsection{Cosine Similarity Convolution for Addressing Varying Amplitudes}
As illustrated in the work of \citeauthor{lhermitte2011comparison} \shortcite{lhermitte2011comparison}, several feature extraction methodologies are highly vulnerable to amplitude changes in time-series recordings. Since dot products in neural networks are vulnerable to amplitude changes \cite{luo2018cosine}, mere application of dot products for bio-medical data might degrade the correctness of identified template signals.

\textbf{Proposed Approach}. To handle this issue, the proposed method extends a general 1D convolutional layer by substituting dot products between convolutional filters and input segments with cosine similarity calculations. On such architecture, output feature map can be calculated as:
\begin{equation}
    o_{i}^{k} = \frac{\mathbf{w}^{k} \cdot \mathbf{x}_{[i;i+L-1]}}{\vert \mathbf{w}^{k} \vert \vert \mathbf{x}_{[i;i+L-1]} \vert}. 
\end{equation}
Furthermore, using the derivatives of cosine similarity in the previous work \cite{luo2018cosine}, the derivative of error with regard to positions in convolutional filter weights can be written as:
\begin{align}
    \frac{\partial E}{\partial w_{i}^{k}} &= \sum_{j}{\frac{\partial E}{\partial o_{j}^{k}}}\bigg( \frac{x_j}{\vert \mathbf{w}^{k} \vert \vert \mathbf{x}_{[j;j+l-1]} \vert} - \frac{w_{i}^{k}(\mathbf{w}^{k} \cdot \mathbf{x}_{[j;j+l-1]})}{\vert \mathbf{w}^{k} \vert^{3} \vert \mathbf{x}_{[j;j+l-1]} \vert} \bigg) \label{eq:backprop1}.
\end{align}
To simplify the calculation, norms of each convolutional filter are normalized to one in our implementation. Thus, resulting forward and backward operation in cosine similarity convolution can be simplified as:
\begin{align}
    o_{i}^{k} &= \frac{\mathbf{w}^{k} \cdot \mathbf{x}_{[i;i+L-1]}}{\vert \mathbf{x}_{[i;i+L-1]} \vert} \label{eq:approx_f} 
\end{align}

\begin{align}
    \frac{\partial E}{\partial w_{i}^{k}} &= \sum_{j}{\frac{\partial E}{\partial o_{j}^{k}}}\frac{x_{i+j-1}}{\vert \mathbf{x}_{[j;j+l-1]} \vert} \label{eq:approx_b}.
\end{align}

With 1-max pooling layer introduced before, gradient values in the proposed method are further simplified as 
\begin{align}
    \frac{\partial E}{\partial w_{i}^{k}} &= \frac{\partial E}{\partial o_{j^{*}}^{k}}\frac{x_{i+j^{*}-1}}{\vert \mathbf{x}_{[j^{*};j^{*}+l-1]} \vert}.
\end{align}
In the above equation, gradient values will only reflect the waveform shapes of the significant segments regardless of amplitude changes in data. Thus, neural networks will better locate the template patterns in data and update their parameters from them.

\subsection{Pre-training Using Template Patterns}
With above-mentioned two components, cosine similarity convolutional layer and 1-max pooling architecture, each convolutional filter will learn desired template patterns from data. Weights in these filters will be used in initializing the first layer of convolutional filters of target architectures. With template patterns acquired with pre-training, models will more easily locate significant features, which are of more distinguishable and interpretable shapes, and learn to utilize those signals in classifying EEG recordings.


\section{Experimental Setting}
\label{sec: exp_setting}
\subsection{Dataset and Pre-processing}
Models were evaluated on datasets from the previous sleep studies \cite{supratak2017Deepsleepnet,phan2019seqsleepnet,perslev2019u}. For all datasets, sleep stages were labeled according to American Academy of Sleep Medicine (AASM), one of the world-standard sleep scoring systems \cite{berry2012rules}. Following preceding works \cite{supratak2017Deepsleepnet,perslev2019u}, we only considered five classes (Wake, N1, N2, N3, REM) that are closely related to sleep stages. Except for Sleep-EDF datasets, we re-scaled EEG recordings to have interquartile range (Q3 - Q1) of 1 and median of 0 \cite{perslev2019u}. For Sleep-EDF dataset, we used raw recordings to reproduce previous results \cite{supratak2017Deepsleepnet,mousavi2019sleepeegnet}. Unlike the previous work \cite{perslev2019u}, however, we did not zero out extreme values in data. Summaries on employed datasets are as follows.

\begin{itemize}
\item{\textbf{Sleep-EDF Dataset} \cite{goldberger2000physiobank,kemp2000analysis} is a representative database for sleep studies. Fpz-Cz channel from healthy subjects was used in experiments. Two versions of the dataset were used: EDF-2013 (39 recordings) and EDF-2018 (153 recordings).}
\item{\textbf{ISRUC-Sleep Dataset} \cite{khalighi2016isruc} is provided from Hospital of Coimbra University. We experimented on the first (100 subjects) and the third group (10 subjects). C4-A1 channel was used.}
\item{\textbf{CAP Dataset} \cite{terzano2002atlas} is a sleep EEG dataset containing 108 recordings. C4-A1 or C3-A2 channel was used.}
\item{\textbf{SVUH-UCD Dataset} \cite{goldberger2000physiobank} is a set of recordings from 25 subjects provided by St. Vincent's University Hospital and University College Dublin. EEG channel C3-A2 was used and two recordings with unknown labels were excluded.}
\end{itemize}

\subsection{Neural Network Architectures}
We experimented on several architectures to demonstrate that our method can be generally applied to sleep staging. In implementing the architectures that will be pre-trained by the proposed method, we slightly modified the structures when first convolutional layer has large strides. For these architectures, striding sizes were changed to 1, since templates cannot be utilized with large strides. Instead, window sizes of max pooling layer that comes right after the first convolutional layer were increased by the same factor. Hence, total number of parameters in the architecture is unchanged and we believe fairness of comparisons is still valid. We give summaries of architectures as follows.

\begin{itemize}
\item{\textbf{Supratak - LSTM} (DeepSleepNet) \cite{supratak2017Deepsleepnet} is a widely referenced architecture consisting of two-branch convolutional layers with different filter sizes (400 and 50), and Long Short-Term Memory (LSTM) layers \cite{perslev2019u,mousavi2019sleepeegnet}}. 
\item{\textbf{Supratak - CNN} \cite{supratak2017Deepsleepnet} is a classifier on CNN layers of DeepSleepNet proposed for pre-training phase in original work.}
\item{\textbf{Sors} \cite{sors2018convolutional} is a 12-layer CNN architecture for sleep staging. This model was used as a baseline in the recent work \cite{duggal2020rest}.}
\item{\textbf{Dai} \cite{dai2017very} is a set of CNNs for audio classification. It has also been referenced in sleep staging \cite{andreotti2018multichannel}. Among four models, 5-layer and 11-layer architectures were evaluated.}
\end{itemize}

\subsection{Evaluation on Confidence Calibration}
Confidence calibration refers to concordance between model's output confidence and likeliness of correct prediction \cite{guo2017calibration}. From the perspective of confidence calibration, it is highly desirable to get $N \cdot p$ correct predictions from $N$ predictions with confidence probability of $p$. Such models are deemed as well-calibrated models.
We primarily regard maximum values from Softmax Probabilities on model's output as confidence values \cite{guo2017calibration,hendrycks17baseline}.

\begin{table*}[t]
  \caption{Classification Performances on Big Datasets}
  \label{tab:performance_big}
  \centering
  \begin{tabular}{@{}cl@{\hskip 0.1in}ccc@{\hskip 0.05in}ccc@{\hskip 0.05in}cc@{}}
    \toprule
    &                   & \multicolumn{2}{c}{EDF-2018}                               & 
                        & \multicolumn{2}{c}{ISRUC Group1}                           & 
                        & \multicolumn{2}{c}{CAP}                                    \\
                        \cmidrule(lr){3-4} \cmidrule(lr){6-7} \cmidrule(lr){9-10} 
    Models & Methods    & Macro-F1                       & Accuracy                      & 
                        & Macro-F1                       & Accuracy                      & 
                        & Macro-F1                       & Accuracy                      \\
    \midrule[0.7pt]
    \multirow{2}{*}{\shortstack{Supratak \cite{supratak2017Deepsleepnet} \\ LSTM}}
    & \textbf{Template} & \textbf{0.72722} $\pm 0.027$   & \textbf{0.76528} $\pm 0.029$  & 
                        & \textbf{0.74926} $\pm 0.029$   & \textbf{0.76848} $\pm 0.028$  &
                        & \textbf{0.72967} $\pm 0.038$   & \textbf{0.78408} $\pm 0.036$  \\
    & Baseline          & 0.72215 $\pm 0.040$            & 0.75492 $\pm 0.045$            & 
                        & 0.73264 $\pm 0.036$            & 0.74237 $\pm 0.038$            & 
                        & 0.70149 $\pm 0.037$            & 0.74056 $\pm 0.039$            \\
    \midrule[0.4pt]
    \multirow{2}{*}{\shortstack{Supratak \cite{supratak2017Deepsleepnet} \\ CNN}}
    & \textbf{Template} & \textbf{0.70228} $\pm 0.029$   & \textbf{0.75612} $\pm 0.035$   & 
                        & \textbf{0.70601} $\pm 0.022$   & \textbf{0.73242} $\pm 0.025$   & 
                        & \textbf{0.66846} $\pm 0.029$   & \textbf{0.73906} $\pm 0.033$  \\
    & Baseline          & 0.69733 $\pm 0.030$            & 0.74797 $\pm 0.041$            & 
                        & 0.70039 $\pm 0.023$            & 0.71745 $\pm 0.028$            & 
                        & 0.65264 $\pm 0.035$            & 0.71576 $\pm 0.038$            \\
    \midrule[0.4pt]
    \multirow{2}{*}{Sors \cite{sors2018convolutional}}
    & \textbf{Template} & \textbf{0.72166} $\pm 0.030$   & 0.784621 $\pm 0.027$   & 
                        & 0.71772 $\pm 0.024$            & 0.74543 $\pm 0.025$   & 
                        & \textbf{0.68239} $\pm 0.030$   & \textbf{0.76203} $\pm 0.029$  \\
    & Baseline          & 0.72033 $\pm 0.029$            & \textbf{0.78873} $\pm 0.029$   & 
                        & \textbf{0.71927} $\pm 0.023$   & \textbf{0.74655} $\pm 0.023$   & 
                        & 0.68120 $\pm 0.035$            & 0.75911 $\pm 0.033$            \\
    \midrule[0.4pt]
    \multirow{2}{*}{\shortstack{Dai \cite{dai2017very} \\ 5-layer CNN}}
    & \textbf{Template} & \textbf{0.72700} $\pm 0.033$   & \textbf{0.77974} $\pm 0.033$   & 
                        & \textbf{0.71881} $\pm 0.021$   & \textbf{0.74335} $\pm 0.023$   & 
                        & \textbf{0.71096} $\pm 0.034$   & \textbf{0.78023} $\pm 0.033$   \\
    & Baseline          & 0.72290 $\pm 0.030$            & 0.77677 $\pm 0.038$            & 
                        & 0.70772 $\pm 0.021$            & 0.73023 $\pm 0.026$            & 
                        & 0.69471 $\pm 0.039$            & 0.76652 $\pm 0.042$            \\   
    \midrule[0.4pt]
    \multirow{2}{*}{\shortstack{Dai \cite{dai2017very} \\ 11-layer CNN}}
    & \textbf{Template} & \textbf{0.73285} $\pm 0.030$   & \textbf{0.78930} $\pm 0.035$   & 
                        & \textbf{0.72518} $\pm 0.022$   & 0.75200 $\pm 0.025$            & 
                        & \textbf{0.71061} $\pm 0.041$   & \textbf{0.78390} $\pm 0.040$   \\
    & Baseline          & 0.72635 $\pm 0.027$            & 0.78702 $\pm 0.034$            & 
                        & 0.72436 $\pm 0.018$            & \textbf{0.75377} $\pm 0.018$   & 
                        & 0.70671 $\pm 0.035$            & 0.78296 $\pm 0.030$            \\                           
    \bottomrule
  \end{tabular}
\end{table*}

In evaluating the confidence calibration, we used \textbf{Expected Calibration Error (ECE)} \cite{naeini2015obtaining}. This metric approximates miscalibration rates with differences between confidences and actual accuracies of each probabilistic bin. This can be formulated as
\begin{equation}
    \text{ECE} = \sum_{m=1}^{M} \frac{\mid B_m \mid}{n} \Big\vert acc(B_m) - conf(B_m) \Big\vert,
\end{equation}
with M denoting the number of bins. To further inspect the calibration of confidences, we provide \textbf{Reliability Diagram} \cite{niculescu2005predicting}. This diagram visualizes the gap between likeliness of correct predictions and desirable accuracies for each confidence interval. On this diagram, when average accuracy is smaller than the desirable accuracy for some intervals, we can infer that model is over-confident about their predictions. 

\subsection{Interpretation on Model} \label{sec: interpretation on model}
\textbf{Interpretation on Template Patterns} \\ 
To investigate the constructed templates and how they were utilized during prediction, we interpreted learned representations in convolutional filters. Especially, Sleep Spindles and K-complexes are well-known template patterns in EEG recordings observed during sleep stage 2. We employed DETOKS algorithm \cite{parekh2015detection}, which is proposed for detecting Sleep Spindles and K-complexes from raw EEG signals. Even though such detection algorithms are vulnerable to generate false positives and are incomplete in terms of accuracy \cite{parekh2015detection}, we assumed results from DETOKS as groundtruth during inspection to shed light on how models are utilizing important aspects in data. In Section \ref{sec: interpret}, we compared patterns detected by DETOKS algorithm and templates constructed in our method.

\vskip 0.1in
\noindent \textbf{Saliency Map for Interpreting the Predictions} \\
In a perspective of model interpretability, we show which part of EEG recordings models have paid attention to when making a prediction. Saliency Map \cite{simonyan2013deep} calculates gradient value, or saliency value, of model's output with respect to each data point. Intuitively, data points with high saliency values indicate significant parts exploited during prediction, since model output is more severely affected by changes in those regions compared to other regions. As proposed in the original work, to reveal how models utilized features in input recordings, we calculated the derivative of models output prediction with respect to input recording: $\big\vert{\frac{\partial f(\mathbf{x})}{\partial \mathbf{x}}}\big\vert$, where $f(\mathbf{x})$ is a prediction of model.

\subsection{Implementation Details}
Datasets were divided into train, validation and test data with a ratio of 80\%, 10\% and 10\% on subject basis, which means that recording segments from a subject cannot be contained in both of training and test data. We chose subject-based division scheme to avoid information leak between train and test data. All the recordings were annotated and classified for each of 30 second segment. EEG signals were then re-sampled to 100 Hz before being fed into the model. Reported performances in result section were produced from models with the best Macro-F1 score on validation dataset. Models were implemented in Python 3.6.7 and Pytorch 1.4.0 \cite{paszke2017automatic}. We executed experiments on servers equipped with NVIDIA GeForce RTX 2080 Ti GPUs and Intel Xeon(R) Gold 5118 2.30GHz CPUs. 

\section{Results}
\label{sec: res}
\begin{table*}[]
  \caption{Classification Performances on Small Datasets}
  \label{tab:performance_small}
  \centering
  \begin{tabular}{@{}cl@{\hskip 0.1in}ccc@{\hskip 0.05in}ccc@{\hskip 0.05in}cc@{}}
    \toprule
    &                   & \multicolumn{2}{c}{EDF-2013}                               & 
                        & \multicolumn{2}{c}{ISRUC Group3}                           & 
                        & \multicolumn{2}{c}{SVUH-UCD}                               \\
                        \cmidrule(lr){3-4} \cmidrule(lr){6-7} \cmidrule(lr){9-10} 
    Models & Methods    & Macro-F1                       & Accuracy                      & 
                        & Macro-F1                       & Accuracy                      & 
                        & Macro-F1                       & Accuracy                      \\
    \midrule[0.7pt]
    \multirow{2}{*}{\shortstack{Supratak \cite{supratak2017Deepsleepnet} \\ LSTM}}
    & \textbf{Template} & \textbf{0.78152} $\pm 0.051$   & \textbf{0.82728} $\pm 0.052$  & 
                        & \textbf{0.73621} $\pm 0.046$   & \textbf{0.76393} $\pm 0.047$  &
                        & \textbf{0.59814} $\pm 0.102$   & \textbf{0.61415} $\pm 0.114$  \\
    & Baseline          & 0.77524 $\pm 0.047$            & 0.81704 $\pm 0.053$           & 
                        & 0.63802 $\pm 0.087$            & 0.67448 $\pm 0.090$           & 
                        & 0.57672 $\pm 0.084$            & 0.60235 $\pm 0.086$           \\
    \midrule[0.4pt]
    \multirow{2}{*}{\shortstack{Supratak \cite{supratak2017Deepsleepnet} \\ CNN}}
    & \textbf{Template} & \textbf{0.72981} $\pm 0.054$   & \textbf{0.79175} $\pm 0.057$   & 
                        & \textbf{0.68174} $\pm 0.051$   & \textbf{0.73134} $\pm 0.050$   & 
                        & \textbf{0.55026} $\pm 0.051$   & \textbf{0.55946} $\pm 0.056$  \\
    & Baseline          & 0.72513 $\pm 0.049$            & 0.79128 $\pm 0.053$            & 
                        & 0.65717 $\pm 0.051$            & 0.69898 $\pm 0.070$            & 
                        & 0.52884 $\pm 0.073$            & 0.55744 $\pm 0.082$            \\
    \midrule[0.4pt]
    \multirow{2}{*}{Sors \cite{sors2018convolutional}}
    & \textbf{Template} & \textbf{0.74566} $\pm 0.052$   & 0.81823 $\pm 0.049$            & 
                        & \textbf{0.62768} $\pm 0.053$   & \textbf{0.69159} $\pm 0.055$   & 
                        & \textbf{0.45682} $\pm 0.051$   & \textbf{0.49390} $\pm 0.069$  \\
    & Baseline          & 0.74265 $\pm 0.050$            & \textbf{0.81902 $\pm 0.047$}   & 
                        & 0.62057 $\pm 0.051$            & 0.68300 $\pm 0.058$            & 
                        & 0.44484 $\pm 0.052$            & 0.47606 $\pm 0.061$            \\
    \midrule[0.4pt]
    \multirow{2}{*}{\shortstack{Dai \cite{dai2017very} \\ 5-layer CNN}}
    & \textbf{Template} & \textbf{0.76623} $\pm 0.056$   & \textbf{0.81957} $\pm 0.059$   & 
                        & 0.65811 $\pm 0.055$            & 0.71086 $\pm 0.052$            & 
                        & \textbf{0.58210} $\pm 0.074$   & \textbf{0.60416} $\pm 0.094$   \\
    & Baseline          & 0.74841 $\pm 0.060$            & 0.80078 $\pm 0.060$            & 
                        & \textbf{0.67422} $\pm 0.056$   & \textbf{0.72598} $\pm 0.048$   & 
                        & 0.57016 $\pm 0.073$            & 0.58874 $\pm 0.088$   \\  
    \midrule[0.4pt]
    \multirow{2}{*}{\shortstack{Dai \cite{dai2017very} \\ 11-layer CNN}}
    & \textbf{Template} & \textbf{0.75892} $\pm 0.064$   & \textbf{0.81894} $\pm 0.062$   &
                        & \textbf{0.68474} $\pm 0.063$   & \textbf{0.73363} $\pm 0.056$   &
                        & \textbf{0.58110} $\pm 0.086$   & \textbf{0.59897} $\pm 0.098$   \\
    & Baseline          & 0.74894 $\pm 0.056$            & 0.80500 $\pm 0.056$            &
                        & 0.67806 $\pm 0.050$            & 0.72850 $\pm 0.054$            &
                        & 0.57033 $\pm 0.078$            & 0.58527 $\pm 0.094$            \\  
    \bottomrule
  \end{tabular}
\end{table*}

In this section, we refer (1) models trained with the proposed work as \textbf{Template} and (2) models from the original work of target neural networks as \textbf{Baseline}. Results on classification performances and confidence calibration are provided. Afterwards, we estimated robustness in several aspects. Lastly, we visually investigated and interpreted learned representations.

\subsection{Classification Performances}
Table \ref{tab:performance_big} lists the classification performances measured from large datasets containing 100 recordings or more. Throughout several architectures, models trained with the proposed method achieved better scores compared to baseline architectures in terms of Macro-F1 scores and accuracies. For Sors, although our model performed better on EDF-2018 and CAP dataset, differences in scores are not significant compared to other architectures. In ISRUC, the baseline model was even better compared to ours. We believe such result is due to the small filter size employed in the first convolutional layer of Sors. Since the first convolutional layer has filter length of 7 in Sors, it might be harder for the proposed method to extract template patterns with such short filters. Thus, models might fail in fully utilizing the advantages of template pre-training.

Additional experiments were conducted on small datasets. Macro-F1 scores measured from datasets with fewer than 50 subjects are provided in Table \ref{tab:performance_small}. Models trained with the proposed method show larger improvements in classification performances compared to results reported in Table \ref{tab:performance_big}. For example, in ISRUC Group3, Macro-F1 scores for LSTM model are improved by $15\%$ on average by adopting the proposed pre-training, whereas Macro-F1 score increased only by $2\%$ in ISRUC Group1. In Section \ref{sec:robustness}, we thoroughly inspected the robustness of the proposed method to dataset size.

To summarize, pre-trained models outperformed baseline models both in overall accuracy and classification of minor classes. Our results indicate that utilizing template patterns to construct representations of EEG recordings is an effective approach in sleep staging tasks. In Section \ref{sec: interpret}, we further investigated representations from the feature level.

\begin{figure}[!t]
  \centering
  \includegraphics[width=\linewidth]{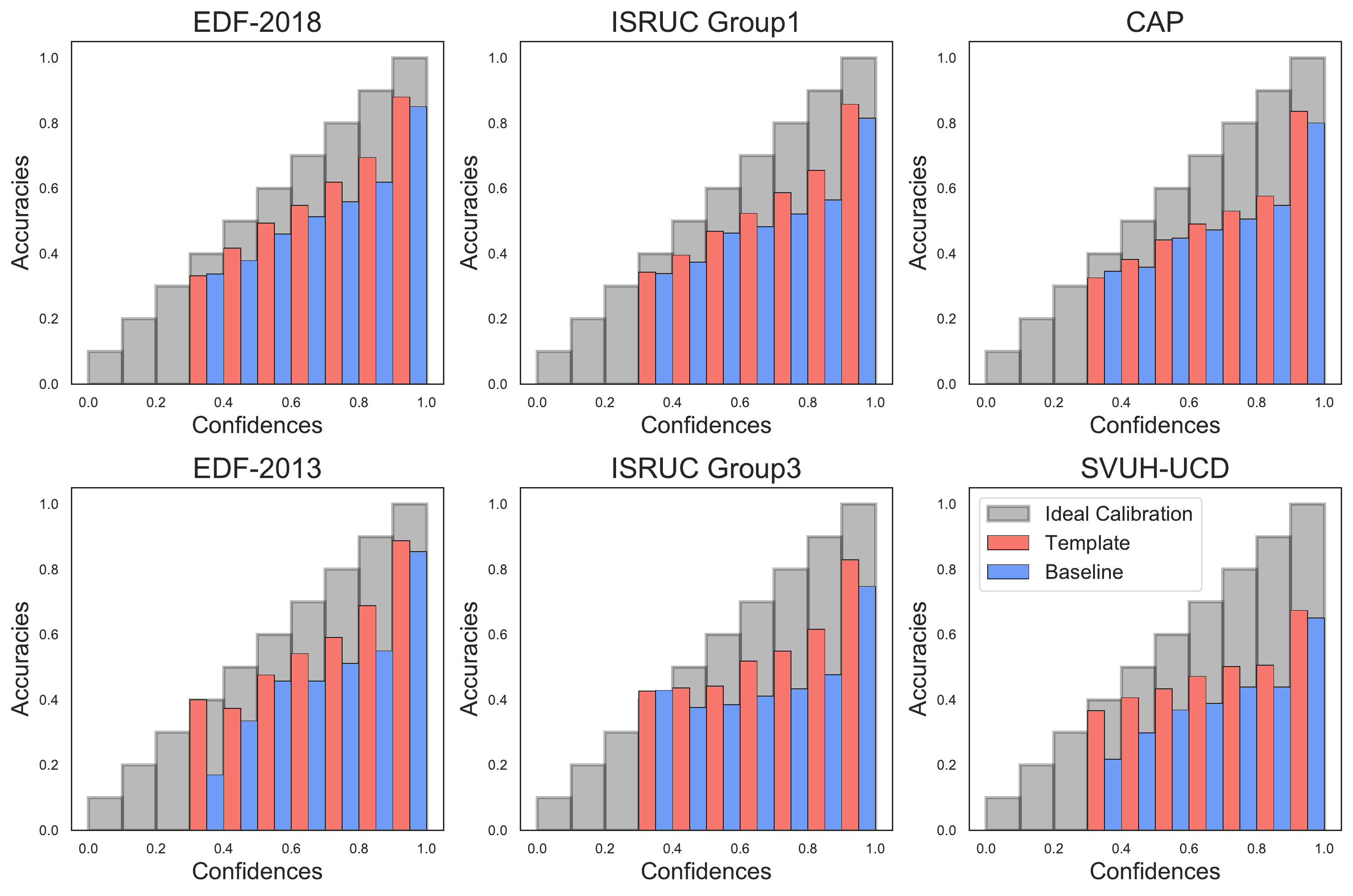}
  \caption{Reliability Diagrams from Six Datasets. Ideal accuracies are drawn as a reference. To avoid misunderstanding, intervals with less than 50 data points are excluded.}
  \Description{Reliability Diagrams}
  \label{fig:reldiagram}
\end{figure}

\subsection{Evaluation on Confidence Calibration} 
We examined the robustness of the proposed method regarding confidences in the output. Similar to previous works \cite{hendrycks17baseline}, maximum values from softmax distribution of model's outputs were interpreted as confidence scores. Table \ref{tab:robustness_confidence} lists ECE scores calculated from the architecture proposed in Supratak's work \cite{supratak2017Deepsleepnet}. For models built on recurrent layers, ECE scores on softmax confidence decreased significantly when TRIER was adopted. Reduced errors on confidence estimation indicate that the proposed method helped models to produce calibrated outputs. Such results are in concordance with the findings that pre-training can enhance robustness regarding model uncertainty \cite{hendrycks19a}. For Sors, confidence calibration was better in ISRUC datasets. However, we believe significant improvements on confidence calibration in Supratak architecture are still meaningful, since modeling a temporal relations between sleep stages with recurrent layers is widely adopted in sleep staging tasks \cite{supratak2017Deepsleepnet,phan2019seqsleepnet,mousavi2019sleepeegnet}.

To further evaluate the degree of calibration on each confidence interval, we generated reliability diagram in Figure \ref{fig:reldiagram}. It can be observed that confidences from the proposed method are overally much more closer to ideal calibration levels. This indicates that our pre-training scheme effectively relieved the over-confident tendency of neural networks \cite{guo2017calibration}.

\begin{table}[t]
  \caption{ECE Scores Calculated for Sleep Datasets}
  \label{tab:robustness_confidence}
  \begin{tabular}{@{}l@{\hskip 0.15in}ccc@{\hskip 0.05in}cc}
    \toprule
                        & \multicolumn{2}{c}{Supratak}                      &
                        & \multicolumn{2}{c}{Sors}                 \\
                        \cmidrule(lr){2-3} \cmidrule(lr){5-6}
    Dataset             & Baseline                      & Template              &  
                        & Baseline                      & Template              \\
    \midrule
    EDF-2013            & 14.53                         & \textbf{10.96}        &
                        & 14.38                         & \textbf{14.21}         \\
    EDF-2018            & 14.45                         & \textbf{11.09}        &
                        & 16.28                         & \textbf{13.91}        \\
    ISRUC Group1        & 17.95                         & \textbf{13.26}        &
                        & \textbf{11.49}                & 13.52                 \\
    ISRUC Group3        & 25.71                         & \textbf{15.32}        &
                        & \textbf{9.68}                & 10.50                 \\
    CAP                 & 19.49                         & \textbf{15.79}        &
                        & 13.73                         & \textbf{12.10}        \\
    SVUH-UCD            & 33.48                         & \textbf{29.49}        &
                        & 21.44                         & \textbf{19.77}        \\
  \bottomrule
\end{tabular}
\end{table}

\begin{table*}[t]
  \caption{Macro-F1 Scores (ratio compared to results from Table \ref{tab:performance_big}) with Varying the Training Size}
  \label{tab:train_size_f1}
  \centering
  \begin{tabular}{@{}c@{\hskip 0.1in}l cccccccc@{}}
    \toprule
    & Data Size         & 80                         & 70                         
                        & 60                         & 50                      
                        & 40                         & 30
                        & 20                         & 10                         \\
    \midrule[0.7pt]
    \multirow{2}{*}{EDF}
    & \textbf{Template} & \textbf{0.725} (1.00)    & \textbf{0.726} (1.00)
                        & \textbf{0.721} (0.99)    & \textbf{0.710} (0.98)
                        & \textbf{0.709} (0.98)    & \textbf{0.696} (0.96)
                        & \textbf{0.677} (0.93)    & \textbf{0.642} (0.88)     \\
    & Baseline          & 0.718 (1.00)             & 0.715 (0.99)
                        & 0.713 (0.99)             & 0.703 (0.97)
                        & 0.699 (0.97)             & 0.679 (0.94)
                        & 0.665 (0.92)             & 0.635 (0.88)             \\
    \midrule[0.4pt]
    \multirow{2}{*}{ISRUC}
    & \textbf{Template} & \textbf{0.750} (1.00)    & \textbf{0.750} (1.00)   
                        & \textbf{0.750} (1.00)    & \textbf{0.745} (0.99)  
                        & \textbf{0.737} (0.98)    & \textbf{0.733} (0.98)  
                        & \textbf{0.720} (0.96)    & \textbf{0.698} (0.93)    \\
    & Baseline          & 0.740 (1.00)             & 0.737 (1.00)             
                        & 0.733 (1.00)             & 0.728 (0.99)
                        & 0.720 (0.98)             & 0.710 (0.97)
                        & 0.690 (0.94)             & 0.664 (0.91)             \\
    \midrule[0.4pt]
    \multirow{2}{*}{CAP}
    & \textbf{Template} & \textbf{0.726} (1.00)    & \textbf{0.725} (0.99)   
                        & \textbf{0.722} (0.99)    & \textbf{0.718} (0.98)  
                        & \textbf{0.709} (0.97)    & \textbf{0.695} (0.95)  
                        & \textbf{0.684} (0.94)    & \textbf{0.655} (0.90)    \\
    & Baseline          & 0.688 (0.98)             & 0.699 (1.00)             
                        & 0.693 (0.99)             & 0.693 (0.99)
                        & 0.684 (0.97)             & 0.674 (0.96)
                        & 0.660 (0.94)             & 0.630 (0.90)             \\
    \bottomrule
  \end{tabular}
\end{table*}

\begin{table*}[t]
  \caption{ECE Scores (ratio compared to results from Table \ref{tab:robustness_confidence}) with Varying the Training Size}
  \label{tab:train_size_ece}
  \centering
  \begin{tabular}{@{}c@{\hskip 0.1in}l cccccccc@{}}
    \toprule
    & Data Size         & 80                         & 70                         
                        & 60                         & 50                      
                        & 40                         & 30
                        & 20                         & 10                         \\
    \midrule[0.7pt]
    \multirow{2}{*}{EDF}
    & \textbf{Template} & \textbf{13.54} (1.22)      & \textbf{14.00} (1.26) 
                        & \textbf{13.49} (1.22)      & \textbf{15.90} (1.43) 
                        & \textbf{16.97} (1.53)      & \textbf{15.96} (1.44)
                        & \textbf{20.25} (1.83)      & \textbf{23.16} (2.09)    \\
    & Baseline          & 16.55 (1.15)               & 17.68 (1.22)
                        & 17.86 (1.24)               & 19.00 (1.31)
                        & 19.90 (1.38)               & 21.39 (1.48)
                        & 23.20 (1.61)               & 24.78 (1.71)             \\
    \midrule[0.4pt]
    \multirow{2}{*}{ISRUC}
    & \textbf{Template} & \textbf{14.00} (1.06)      & \textbf{13.55} (1.02)   
                        & \textbf{14.79} (1.12)      & \textbf{15.89} (1.20)  
                        & \textbf{15.88} (1.20)      & \textbf{15.89} (1.20)  
                        & \textbf{17.56} (1.32)      & \textbf{19.75} (1.49)    \\
    & Baseline          & 16.85 (0.94)               & 17.69 (0.99)            
                        & 18.20 (1.01)               & 19.13 (1.07)
                        & 19.97 (1.11)               & 19.91 (1.11)
                        & 21.79 (1.21)               & 24.24 (1.35)            \\
    \midrule[0.4pt]
    \multirow{2}{*}{CAP}
    & \textbf{Template} & \textbf{13.92} (0.88)      & \textbf{14.41} (0.91)
                        & \textbf{13.74} (0.87)      & \textbf{13.46} (0.85)
                        & \textbf{16.11} (1.02)      & \textbf{16.37} (1.04)
                        & \textbf{17.84} (1.13)      & \textbf{19.69} (1.25) \\
    & Baseline          & 20.41 (1.05)               & 20.24 (1.04)     
                        & 19.96 (1.02)               & 20.10 (1.03)
                        & 21.26 (1.09)               & 22.08 (1.13)
                        & 22.73 (1.17)               & 24.82 (1.27)         \\
    \bottomrule
  \end{tabular}
\end{table*}

\begin{figure*}[] 
  \begin{subfigure}[]{0.495\linewidth}
    \centering
    \includegraphics[width=\linewidth]{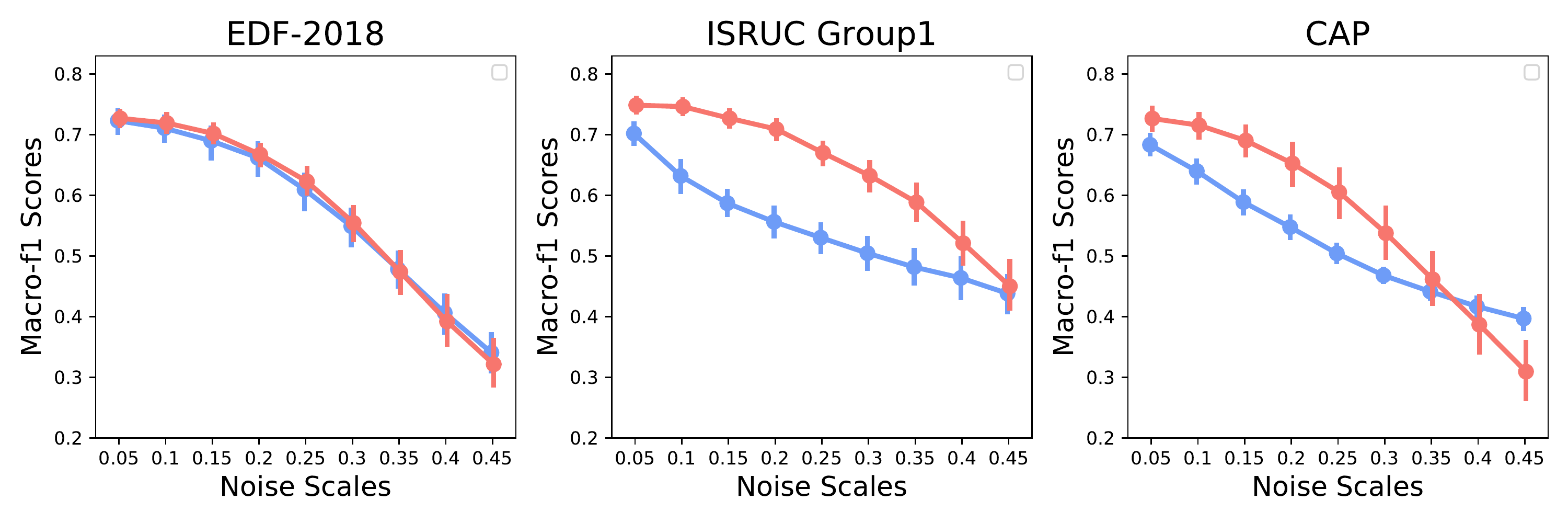}
    \caption{Macro-F1 Scores on Different Noise Scales}
    \label{fig:noise_f1}
  \end{subfigure} \hfill
  \begin{subfigure}[]{0.495\linewidth}
    \centering
    \includegraphics[width=\linewidth]{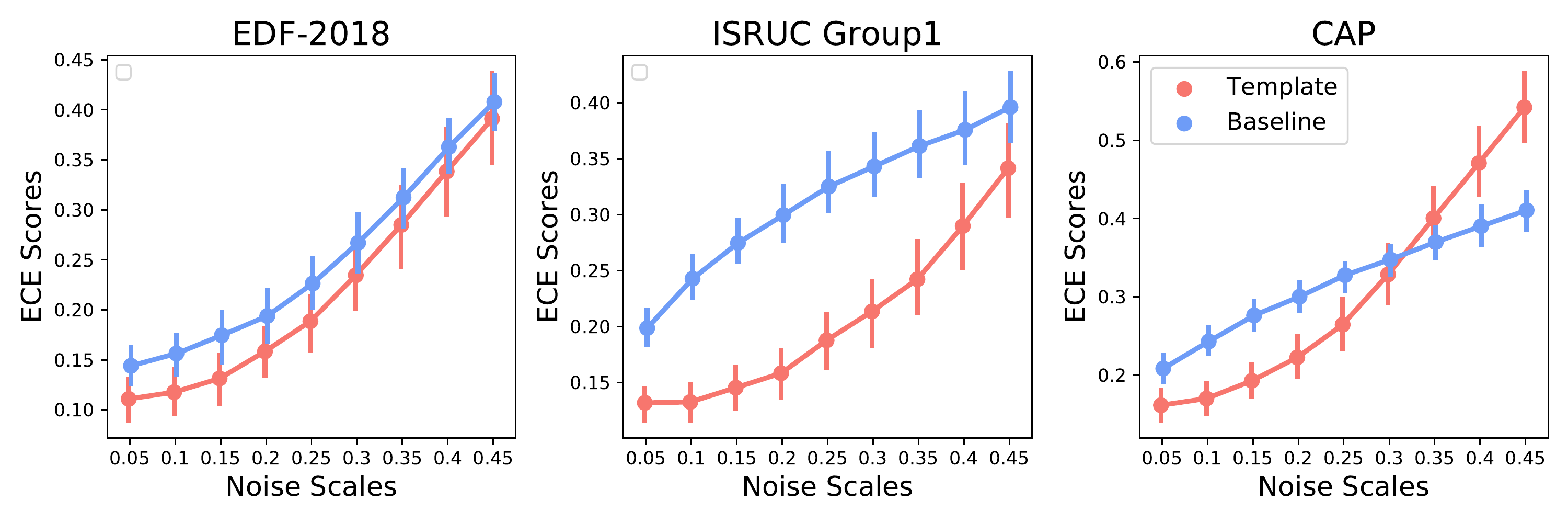}
    \caption{ECE Scores on Different Noise Scales}
    \label{fig:noise_ece}
  \end{subfigure}
\caption{Performances Measured on Gaussian Noises}
\label{fig:noise_robust}
\end{figure*}

\subsection{Evaluation on Robustness} \label{sec:robustness}
\noindent \textbf{Performances with Varying the Dataset Size} \\
To show that the proposed method works out robustly regardless of dataset size, we evaluated performances of models with varying the size of train dataset. Experiments were conducted on datasets with more than 100 subjects: EDF-2018, ISRUC Group1 and CAP dataset. Results are listed in Table \ref{tab:train_size_f1} and Table \ref{tab:train_size_ece}. The proposed method consistently shows better performances throughout various dataset sizes. Similarly, ECE scores are significantly lower. These results indicate that models trained from the proposed strategy are robust under different dataset sizes. This might be owing to the ability that our method can obtain more generalizable features, which are based on template patterns.

\vskip 0.1in
\noindent \textbf{Robustness to Noise} \\
Since bio-physical signals are easily contaminated by noises \cite{ghanem2018investigation}, it is worth investigating the robustness of models against noises. Among various forms of noises, we experimented using gaussian noises, which were extensively considered in researches on model robustness \cite{hong2019mina,duggal2020rest}. Noise scales were selected with respect to standard deviations of EEG recordings. Figure \ref{fig:noise_robust} demonstrates the performances of models in terms of classification accuracy and confidence calibration. Models trained with the proposed method show better performances with little decrease in Macro-F1 scores until the noise scale reaches $0.3$. Even though Macro-F1 scores do not show significant differences in EDF dataset, performances are consistently higher than the baseline models. Figure \ref{fig:noise_ece} shows ECE scores on various noise levels. As shown in the figure, ECE scores from the proposed method begin to increase at higher noise scales, whereas confidence calibration deteriorates right after the noises are injected. These results demonstrate that owing to template patterns constructed during pre-training, our models succeed in reliably predicting sleep stages from contaminated EEG recordings with better classification performances.

\begin{figure*}[!]
    \centering
    \begin{subfigure}[]{0.495\linewidth}
        \centering
        \includegraphics[width=\linewidth]{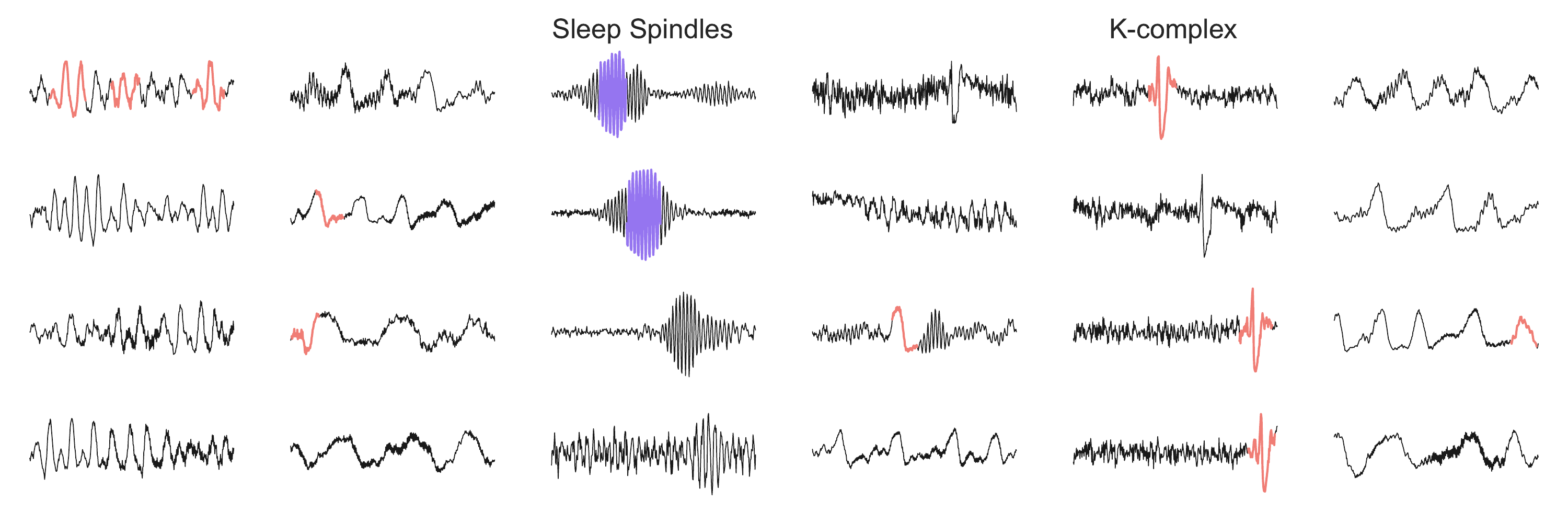}
        \caption{Filters from Template}
        \label{fig:filters_template}
    \end{subfigure} \hfill
    \begin{subfigure}[]{0.495\linewidth}
        \centering
        \includegraphics[width=\linewidth]{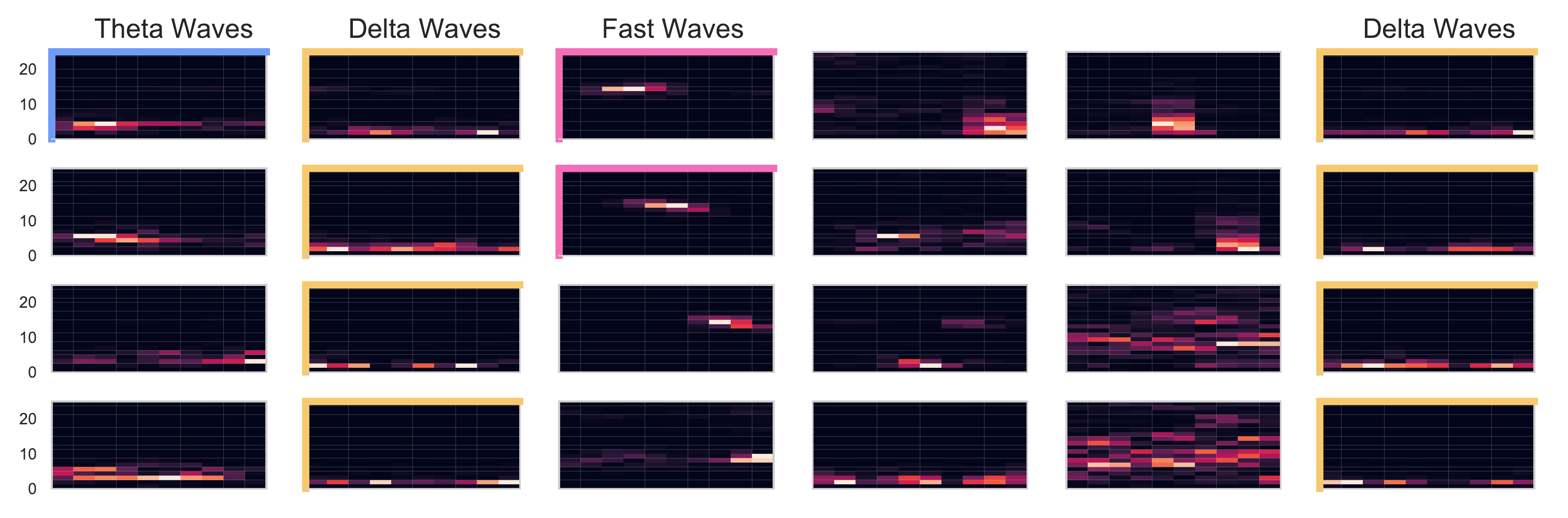}
        \caption{Spectrograms from Template}
        \label{fig:spectrograms_template}
    \end{subfigure} 
    \begin{subfigure}[]{0.495\linewidth}
        \centering
        \includegraphics[width=\linewidth]{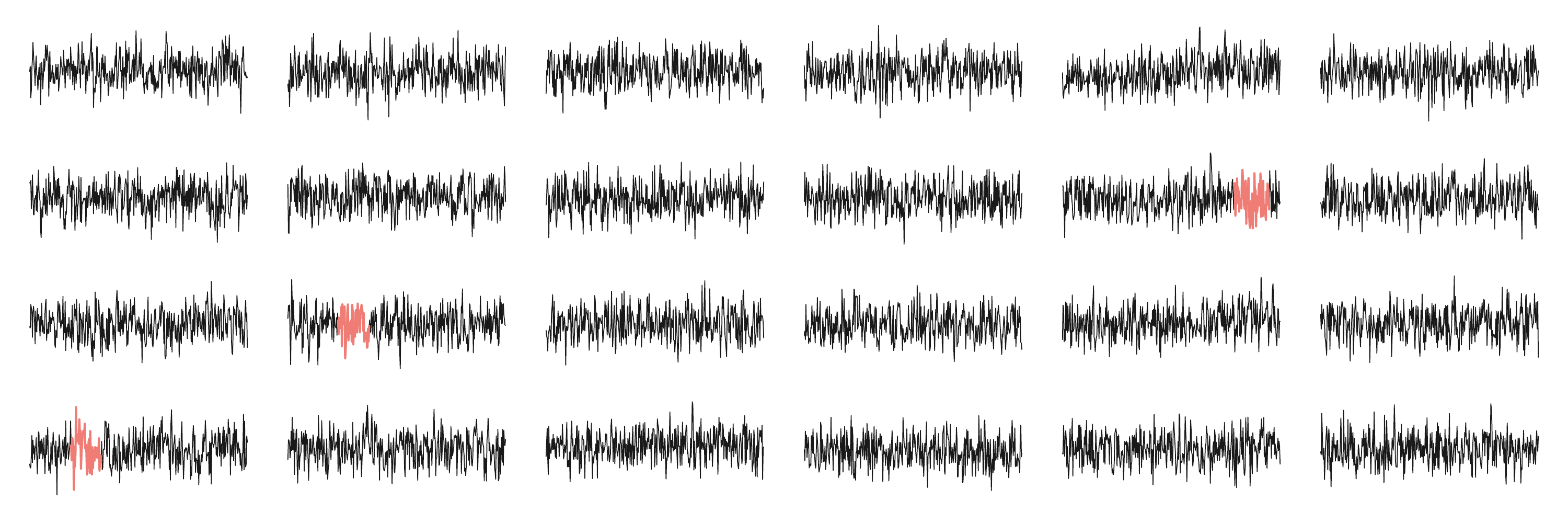}
        \caption{Filters from Baseline}
        \label{fig:filters_baseline}
    \end{subfigure} \hfill
    \begin{subfigure}[]{0.495\linewidth}
        \centering
        \includegraphics[width=\linewidth]{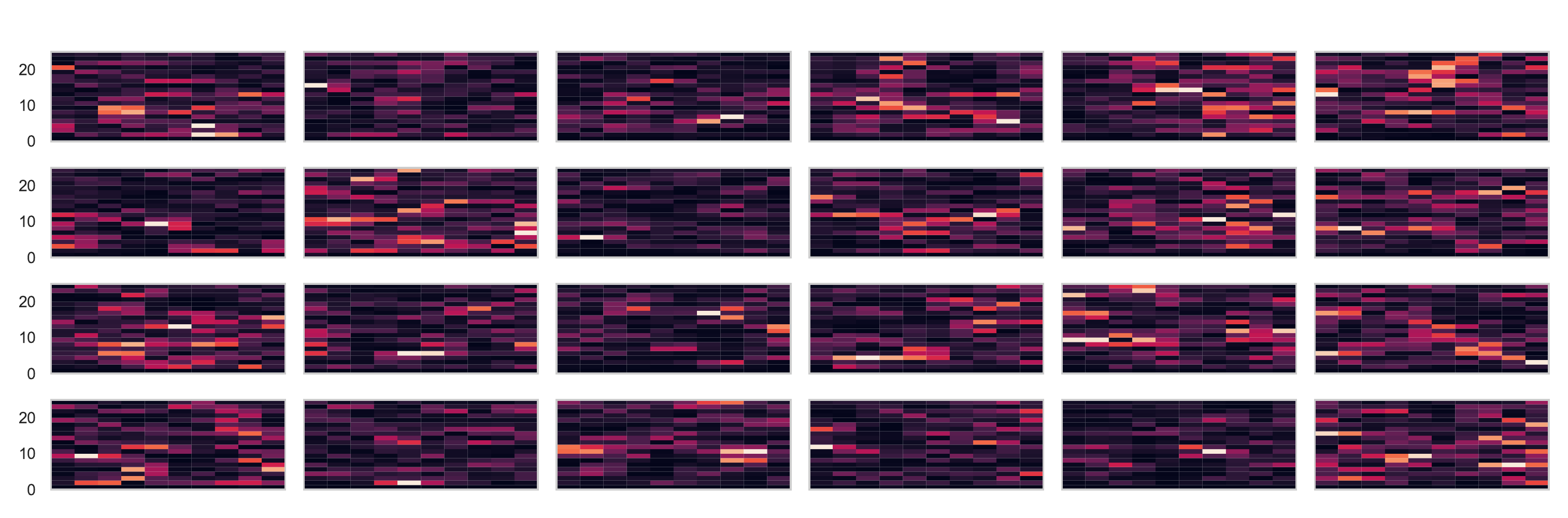}
        \caption{Spectrograms from Baseline}
        \label{fig:spectrograms_baseline}
    \end{subfigure} 
    \caption{Overall Shapes of Convolutional Filters from Supratak \cite{supratak2017Deepsleepnet}. Filters from the proposed method and Baseline are drawn. Regions corresponding to Sleep Spindles and K-complexes detected by DETOKS are highlighted by different colors. On spectrogram, filters that are dedicated to certain frequency bands are marked with boundary colors.}
    \label{fig:filter_cluster}
\end{figure*}

\begin{figure*}[!]
    \centering
    \begin{subfigure}[]{0.495\linewidth}
        \centering
        \includegraphics[width=\linewidth]{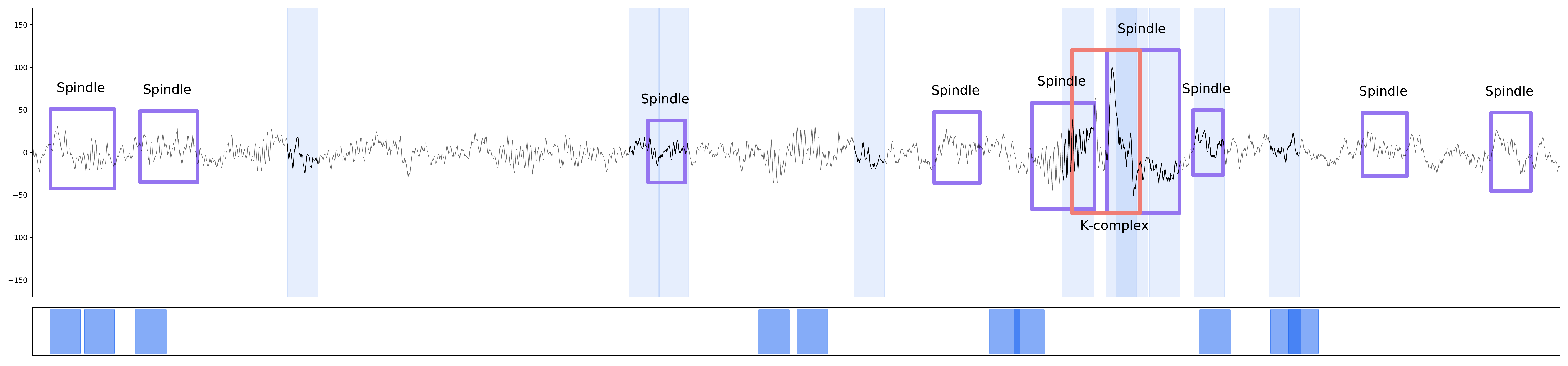}
        \caption{ }
        \label{fig:interpret1}
    \end{subfigure} \hfill
    \begin{subfigure}[]{0.495\linewidth}
        \centering
        \includegraphics[width=\linewidth]{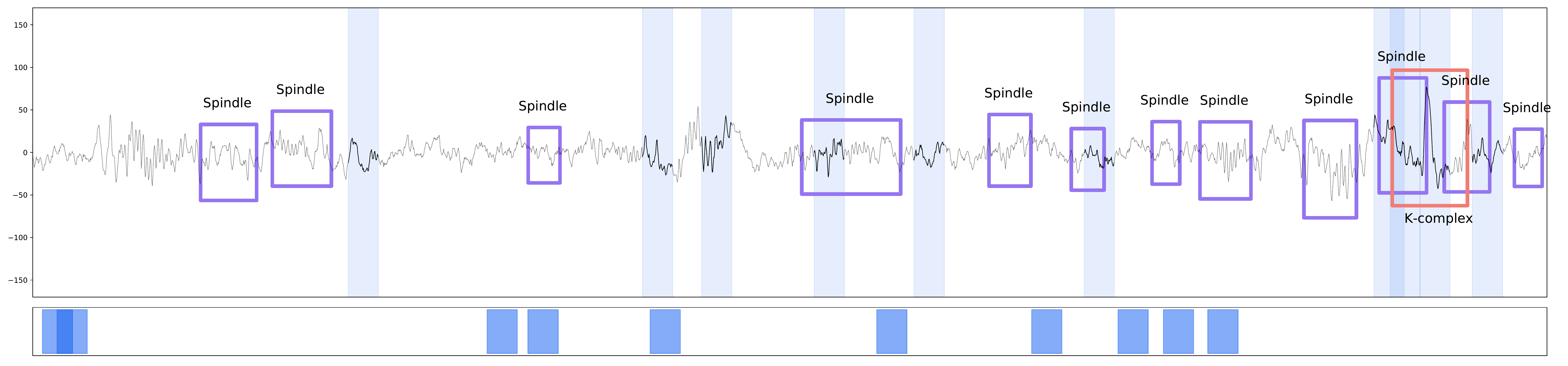}
        \caption{ }
        \label{fig:interpret2}
    \end{subfigure} 
    \begin{subfigure}[]{0.495\linewidth}
        \centering
        \includegraphics[width=\linewidth]{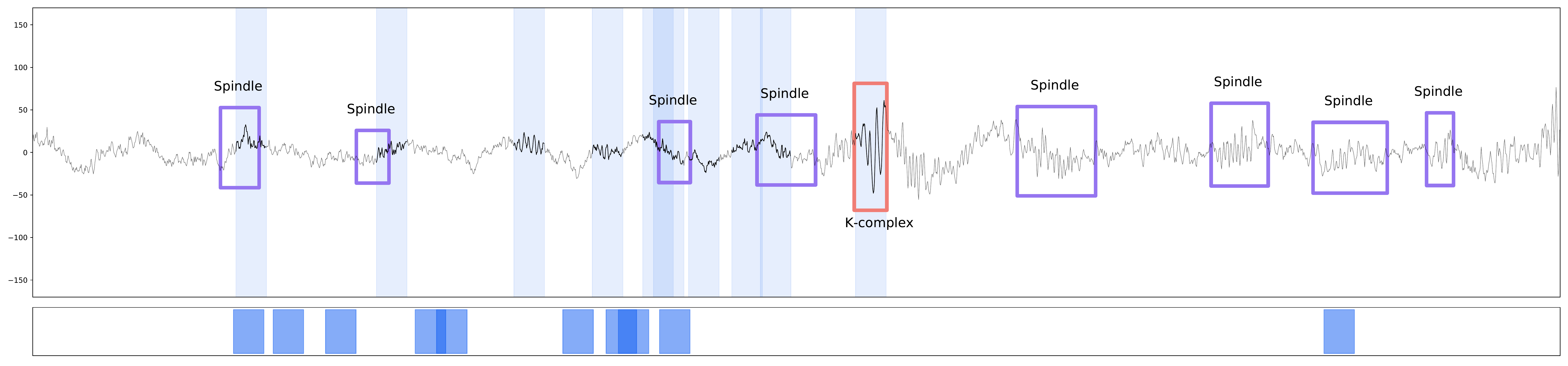}
        \caption{ }
        \label{fig:interpret3}
    \end{subfigure} \hfill
    \begin{subfigure}[]{0.495\linewidth}
        \centering
        \includegraphics[width=\linewidth]{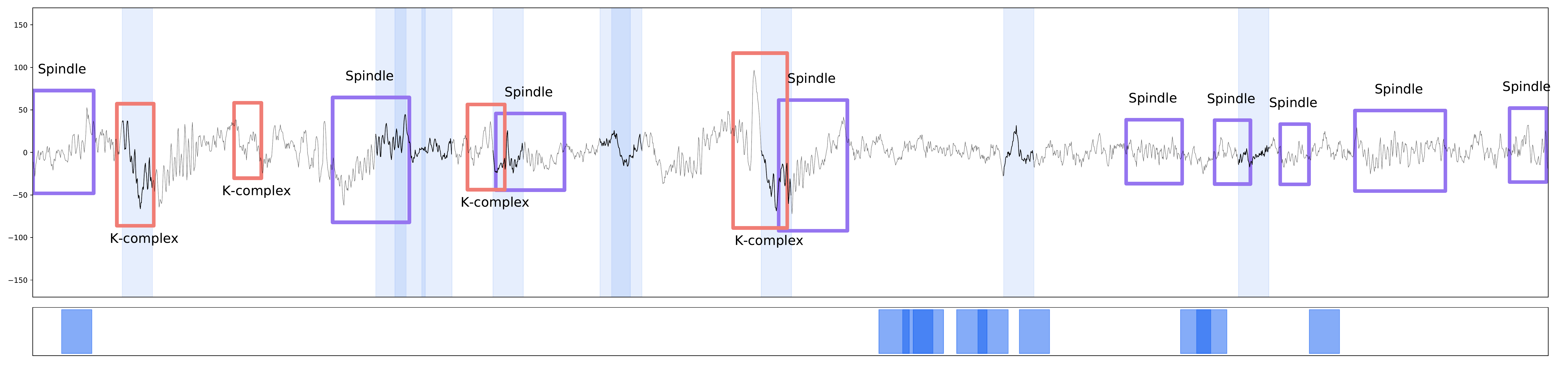}
        \caption{ }
        \label{fig:interpret4}
    \end{subfigure}
    \caption{Interpretations on Supratak \cite{supratak2017Deepsleepnet} When Predicting N2 Stage. In sub-figures, EEG recordings from N2 stage are plotted. Regions corresponding to ten highest saliency values were highlighted. Highlights in the first row are drawn with the saliency map from the proposed method and the second row corresponds to that of Baseline. In addition, K-complex and Sleep Spindles captured by DETOKS are marked with boundary boxes.}
    \label{fig:interpret}
\end{figure*}

\begin{figure*}[!]
    \centering
    \begin{subfigure}[]{0.495\linewidth}
        \centering
        \includegraphics[width=\linewidth]{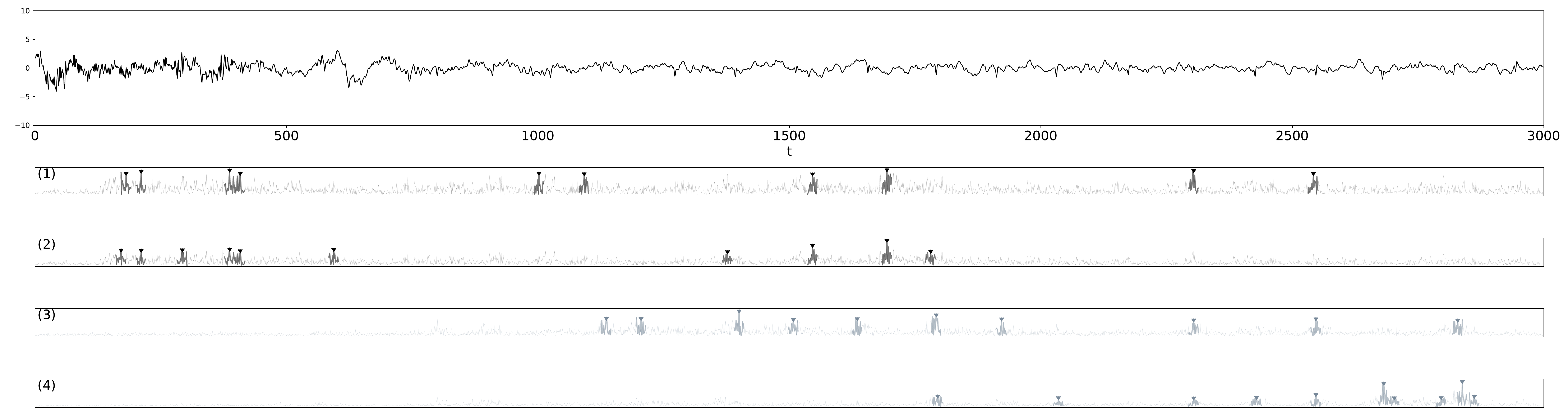}
        \caption{N1 recording}
        \label{fig:noise_interpret1}
    \end{subfigure} \hfill
    \begin{subfigure}[]{0.495\linewidth}
        \centering
        \includegraphics[width=\linewidth]{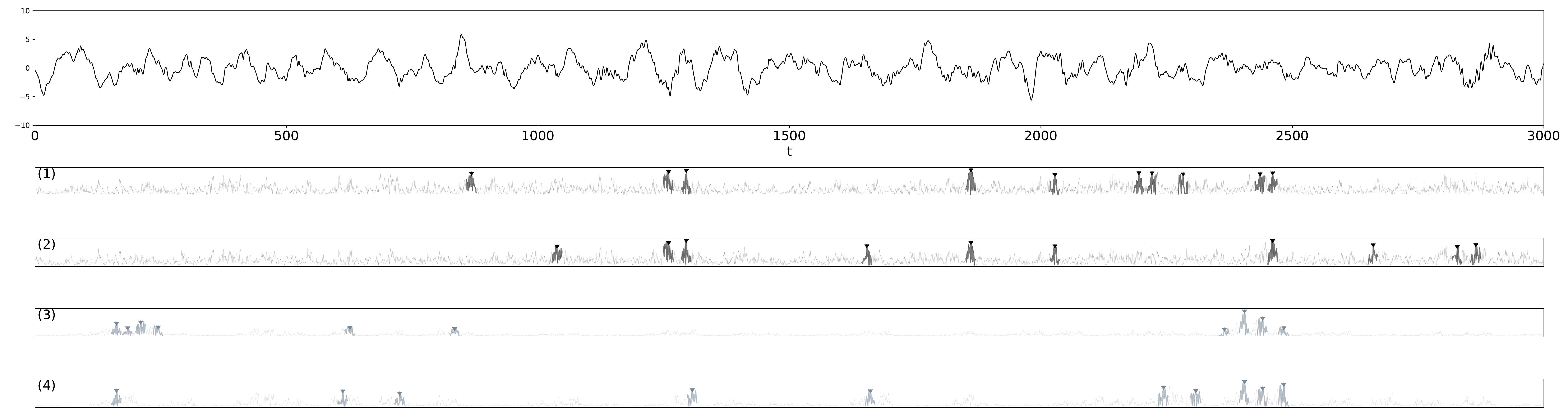}
        \caption{N3 recording}
        \label{fig:noise_interpret2}
    \end{subfigure} 
    \begin{subfigure}[]{0.495\linewidth}
        \centering
        \includegraphics[width=\linewidth]{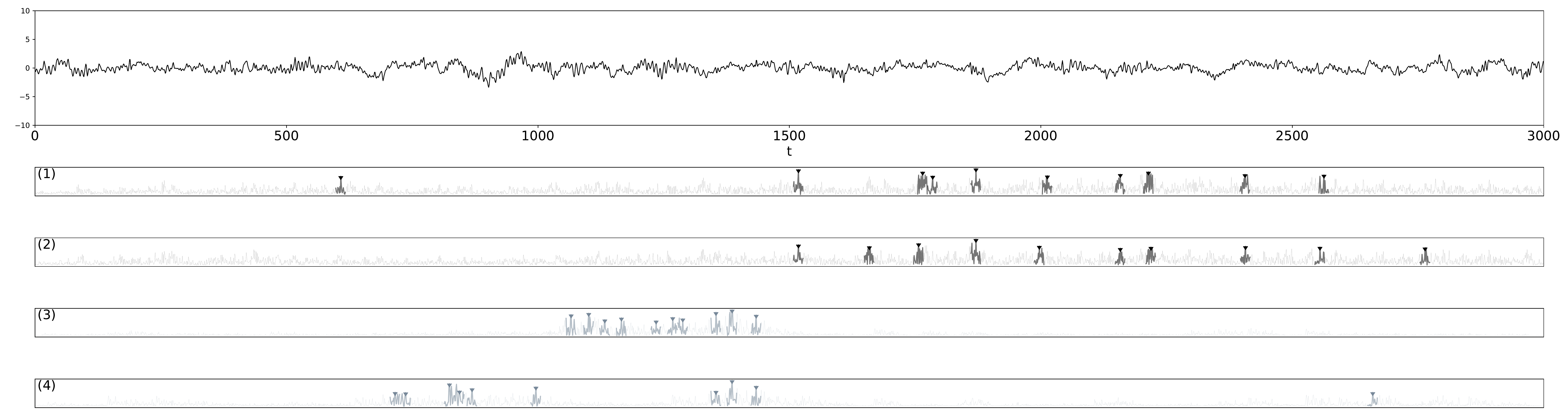}
        \caption{N2 recording}
        \label{fig:noise_interpret3}
    \end{subfigure} \hfill
    \begin{subfigure}[]{0.495\linewidth}
        \centering
        \includegraphics[width=\linewidth]{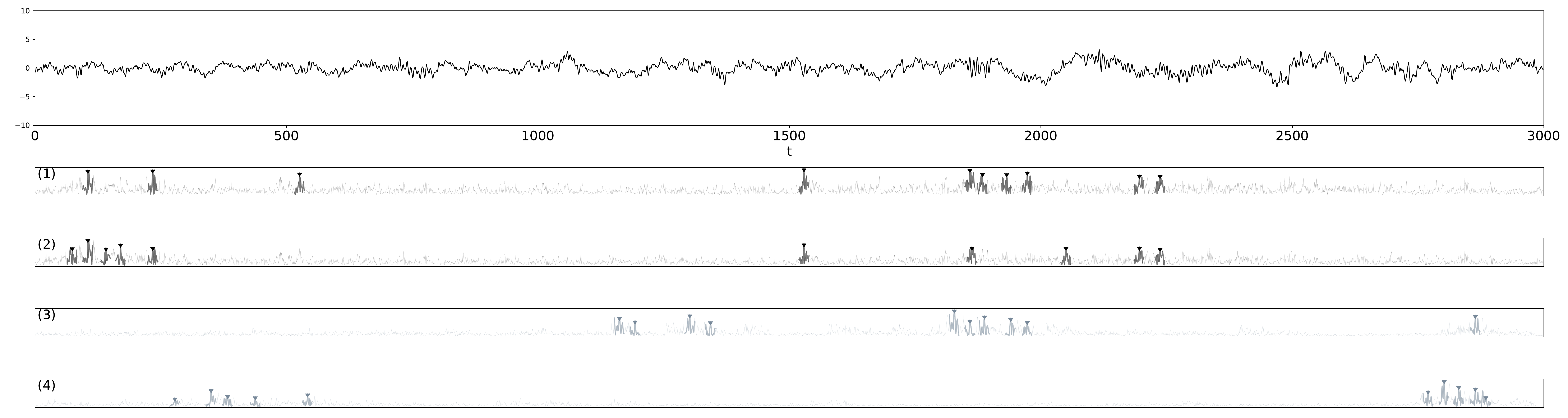}
        \caption{N2 recording}
        \label{fig:noise_interpret4}
    \end{subfigure}
    \caption{Interpretations on Supratak \cite{supratak2017Deepsleepnet} under Gaussian Noises. In sub-figures, EEG recordings are plotted. Under the recordings, four plots of saliency values are visualized with respect to each recording: (1) saliency map on raw data from the proposed method, (2) saliency map on noisy data from the proposed method, (3) saliency map on raw data from  Baseline and (4) saliency map on noisy data from Baseline. Furthermore, data points of ten highest values are marked with \textit{caret down} shape. Scale of noises were configured with ratio of 0.1 from the standard deviation.}
    \label{fig:noise_interpret}
\end{figure*}

\subsection{Investigation and Interpretation on Model} \label{sec: interpret}
\noindent \textbf{Analysis on Learned Filters} \\
To demonstrate that the proposed method effectively learned template patterns, filters in the first convolutional layer of the employed architectures were extensively inspected. We investigated whether convolutional filters correspond to well-known patterns in sleep staging. We applied DETOKS algorithm \cite{parekh2015detection} to the filters. As shown in Figure \ref{fig:filter_cluster}, only the filters from the proposed method correctly contain significant patterns like Sleep Spindles and K-complexes. Even though some false positives were detected as K-complexes by the algorithm, filters from the proposed model obviously contain abrupt burst of signals, which is a characteristic property of K-complex. We further inspected frequency components of each filter. As seen in the figure, filters from the proposed method are dedicated to specific frequency ranges, whereas filters from the baseline model are intermingled by various frequency bands. We marked their spectrograms with boundaries of different colors for each frequency band. As shown in Figure \ref{fig:filter_cluster}, models from the proposed method successfully learned templates of theta waves, delta waves and fast waves.

\vskip 0.1in
\noindent \textbf{Interpretation Based-on Template Patterns} \\
As Sleep Stage N2 contains characteristic EEG patterns, Sleep Spindles and K-complexes, that are only observed during the stage with distinctive shapes, we mainly interpreted EEG segments corresponding to N2 stage. We conducted interpretation experiments focusing on whether models effectively located Sleep Spindles and K-complexes in recordings. We believe such interpretation experiments are meaningful, since experts also utilize presence of these patterns when annotating sleep stages for EEG data \cite{malhotra2013sleep, vsuvsmakova2004human}. In Figure \ref{fig:interpret}, we draw raw EEG signals with highlights on sub-series which have high saliency values. Furthermore, we annotated regions corresponding to Sleep Spindles and K-complexes that were detected by DETOKS \cite{parekh2015detection}. Comparing the saliency maps from the proposed method and the baseline architecture, both models effectively located Sleep Spindles in data, which can be characterized by 12-15 Hz bursts. However, the baseline model did not capture features corresponding to K-complexes, whereas models trained by the proposed method successfully capture K-complexes. We believe such differences are owing to the template filters that resemble K-complexes inspected in the previous section. These results indicate that our method successfully reconstructed template patterns that are referenced by trained experts and models trained with the proposed method can deal with well-known patterns corresponding to each sleep stage.

\vskip 0.1in
\noindent \textbf{Visualizing Robustness to Noises}\\
To investigate why our method is more robust to gaussian noises, we further visualized the changes in saliency values when noises are injected to original data. We believe we can gather understandings on how models respond to noises by inspecting such changes. For this reason, we selected EEG segments that were correctly predicted from both of the proposed method and the baseline model, but mis-classified by the baseline model when noises were injected. In Figure \ref{fig:noise_interpret}, as well as raw recordings, we draw four saliency maps from the proposed method and the baseline architecture with respect to two data: (1) original and (2) noise-contaminated recordings. As shown in the figure, saliency values from the proposed method were relatively consistent under noises. For example, in Figure \ref{fig:noise_interpret1}, we can see that high-saliency regions in the front part ($t < 1000$) and middle part ($1300 < t < 1800$) remain unchanged for the proposed method even with noises. In contrast, for the baseline model, saliency values are largely affected by noises with high-saliency regions in the middle part ($1300 < t < 1800$) mostly decrease. Likewise, in Figure \ref{fig:noise_interpret3}, high-saliency regions in middle part ($1000 < t < 1300$) shift toward front in the baseline model. For Figure \ref{fig:noise_interpret4}, saliency values between 1500 and 1800 have been nearly zeroed out in the baseline architecture. In summary, it can be observed that overall saliency maps are relatively unchanged in the proposed method whereas there are significant changes in saliency values for the baseline model. These observations suggest that feature space remains stable under noises when representations are built on template patterns.


\section{Conclusion}
\label{sec: conclustion}
In this work, we propose a pre-training strategy for neural networks on sleep staging that enables models to utilize template patterns in EEG recordings. We propose a novel neural network, where cosine similarity is introduced as basic operator in order to effectively extract template signals based on significant waveform shapes. Although use of cosine similarity in neural networks has already been introduced in the previous work \cite{luo2018cosine}, as long as we know, this is the first work to utilize cosine similarity to better extract meaningful representations considering the irregular characteristics of data. Extracted patterns were used in initializing the first convolutional layer of target models. With extensive experiments, we revealed that the proposed method works generally well on sleep staging tasks. As well as classification accuracy, overall capacity of the network, including confidence calibration, noise robustness and robustness against small dataset size, was improved by significant degree. Thorough inspection on learned representations and model interpretations demonstrated that models trained with the proposed method were able to utilize important aspects in data that were not handled by the baseline architecture. In summary, template-based representations constructed with the proposed method successfully augmented models in sleep staging tasks in several aspects.




\bibliographystyle{ACM-Reference-Format}
\bibliography{Sections/reference}



\end{document}